\documentclass[journal=jpc, manuscript=article]{achemso}
\usepackage{xr-hyper}
\usepackage[version=3]{mhchem} 
\usepackage{subfig}
\usepackage{threeparttable} 
\usepackage{graphicx}
\usepackage{braket}
\usepackage{xcolor}
\usepackage{soul}
\setstcolor{red}
\usepackage{ulem}

\SectionNumbersOn

\definecolor{airforceblue}{rgb}{0.36, 0.54, 0.66}

\author{\normalsize{Mohammad Aarabi}}
\affiliation{Dipartimento di Chimica Industriale ``Toso Montanari'', Universit{\`a} di Bologna - Alma Mater Studiorum, Via Piero Gobetti 85, 40129 Bologna, Italy}
\alsoaffiliation{Consiglio Nazionale delle Ricerche, Istituto di Chimica dei Composti Organo Metallici (ICCOM-CNR), I-56124 Pisa, Italy}
\author{Emanuele Marsili}
\affiliation{Universit\'e Paris-Saclay, CNRS, Institut de Chimie Physique UMR8000, 91405, Orsay, France}
\author{Massimo Olivucci}
\affiliation{Department of Biotechnology, Chemistry and Pharmacy, Universit\`a degli Studi di Siena, Via A. Moro 2, I-53100 Siena, Italy}
\altaffiliation{Department of Chemistry, Bowling Green State University, Bowling Green, Ohio 43403, United States of America}
\author{David Lauvergnat}
\affiliation{Universit\'e Paris-Saclay, CNRS, Institut de Chimie Physique UMR8000, 91405, Orsay, France}
\author{Federica Agostini}
\affiliation{Universit\'e Paris-Saclay, CNRS, Institut de Chimie Physique UMR8000, 91405, Orsay, France}
\email{federica.agostini@universite-paris-saclay.fr}
\author{Marco Garavelli}
\affiliation{Dipartimento di Chimica Industriale ``Toso Montanari'', Universit{\`a} di Bologna - Alma Mater Studiorum, Via Piero Gobetti 85, 40129 Bologna, Italy}
\email{marco.garavelli@unibo.it}
\author{Fabrizio Santoro}
\affiliation{Consiglio Nazionale delle Ricerche, Istituto di Chimica dei Composti Organo Metallici (ICCOM-CNR), I-56124 Pisa, Italy}
\email{fabrizio.santoro@pi.iccom.cnr.it}

\title[]{Quantum Dynamics Predicts Coherent Oscillatory Behavior in the Early-times of a Photoisomerization Reaction}

\begin{document}
\clearpage
\begin{abstract}
In this work, we study the quantum dynamics of a photoisomerization reaction employing a two-electronic-state three-vibrational-mode model of the 2-\textit{cis}-penta-2,4-dieniminium cation (\textit{cis}-PSB3). In particular, we address two main issues: the challenges encountered in properly converging quantum dynamics calculations, even when a reduced-dimensionality molecular model is used; the emergence of a coherent oscillatory behavior in the formation of the trans isomer upon photoexcitation of \textit{cis}-PSB3. The two issues are strictly related, since only upon reliable convergence, the simulated dynamics is able to capture the large amplitude motion associated to the torsion around the reactive bond, typical of photoisomerizations, which is due to the large amount of  kinetic energy acquired by the vibrational modes after light excitation.
\end{abstract}

\clearpage
\section{Introduction}
Characterizing the quantum dynamics of a molecular system in the excited states is a challenging task, even when reduced dimensionality models are employed to restrict the available configurational space to well-chosen representative ``reaction coordinates''. Often, the selection of such coordinates is guided by the behavior of the molecule in the Franck-Condon region and by normal modes analysis\cite{hahn2000quantum, Domcke_CPL1988,MatsikaJPCA2015, coonjobeeharry2022mixed}, but clearly (and not surprisingly), such modes fail capturing salient dynamic features when these involve large amplitude motions. This is indeed the case of photoisomerization reactions, which are phenomena abundantly explored in the literature for their multiple applications, ranging from photochromism\cite{kortekaas2019evolution} to photoswitches\cite{villaron2020stiff, beharry2011azobenzene} and for their role in photobiology\cite{Schapiro2011, rad2022spiropyran}.

As recently discussed in the context of benchmarks for excited-state molecular dynamics,\cite{cigrang2025roadmap} multidimensional analytical models and the corresponding quantum dynamics are crucial elements in the development of trajectory-based techniques. Indeed, ultimately, trajectory-based methods are (more) easily employed to access the full configurational space, but their performance is often tested in low dimensions. Nonetheless, the assessment of trajectory-based dynamics relies on the accuracy of quantum dynamics, and it is for this reason that the greatest care is needed when providing reliable benchmarks. 

In this contribution, we discuss the challenges encountered to simulate the quantum dynamics of a photoisomerization reaction focusing on a two-state three-di\-men\-sio\-nal model for the 2-\textit{cis}-penta-2,4-dieniminium cation (\textit{cis}-PSB3). In this \textsl{simple} model, three vibrational modes drive the photoreaction, namely the bond-length-alternation (BLA) stretching, the torsional deformation around the reactive double bond (Tors) and the hydrogen-out-of-plane (HOOP) wagging. The analytical form for the diabatic potential energy surfaces along with the corresponding couplings was developed in Ref.[\citenum{marsili2019two}], based on XMCQDPT2\cite{Granovsky2011} ab initio multi-state multi-configurational electronic-structure calculations. In Ref.[\citenum{marsili2020quantum}], the kinetic energy operator and the geometric tensor associated to the BLA, Tors and HOOP curvilinear coordinates were developed, aiming to employ the model for quantum and quantum-classical trajectory-based dynamics. The dynamics studied in Ref.[\citenum{marsili2020quantum}] using such model potential shows that, upon photoexcitation of \textit{cis}-PSB3 from the electronic ground state S$_0$ to the first excited state S$_1$, the activation of the BLA vibrations initially drives the system along the Tors coordinate to access the trans isomer (\textit{trans}-PSB3), also guided by the coupling to the HOOP mode. In the present work we demonstrate that in Ref.[\citenum{marsili2020quantum}] the exact nature of the large amplitude motion along the Tors coordinate was not fully captured due to the limited basis set employed in the quantum dynamics calculations. A related in-depth analysis of the problem motivated us to perform additional simulations that allowed to uncover a qualitatively different time evolution of the cis and trans isomer populations, and thus of the photoisomerization quantum yield. We found that these simulations closely agree with some of the trajectory-based quantum-classical results reported in Ref.[\citenum{marsili2020quantum}].

Uncovering and addressing the challenges encountered to capture quantum dynamical large amplitude motions typical of photoisomerizations led us to identify exciting features during the relaxation dynamics of the photoexcited \textit{cis}-PSB3 to the ground state. In particular, as we will present below, we clearly observe an early-time oscillatory behavior in the reconstitution of the \textit{cis}-PSB3 ground state population which starts just after 50~fs. These oscillations can be associated to the activation of the HOOP wagging, confirming the key role of HOOP in shaping the relaxation dynamics. Furthermore, at later times, a second oscillatory feature corresponding to the ground state ``back-reaction'' from the \textit{trans}-PSB3 product to the \textit{cis}-PSB3 reactant is clearly captured in the behavior of the population of the trans isomer. Both oscillatory features are observed in trajectory-based results but were missing in the original quantum dynamics~\cite{marsili2020quantum}. 

The remainder of the paper is devoted, first, to briefly describe the model potentials and to provide information about the quantum dynamics simulations (Section~\ref{sec: model}), and, second, to analyze the above-mentioned coherent oscillatory features (Section~\ref{sec: results}). Our conclusions are stated in Section~\ref{sec: conclusions}, aiming to raise awareness in the community towards the challenging nature of photoisomerization reactions when described quantum mechanically using reduced dimensionality models.

\section{Presentation of the model and computational details}\label{sec: model}

\begin{figure*}[ht!]
    \centering
    \includegraphics[width=1.0\textwidth]{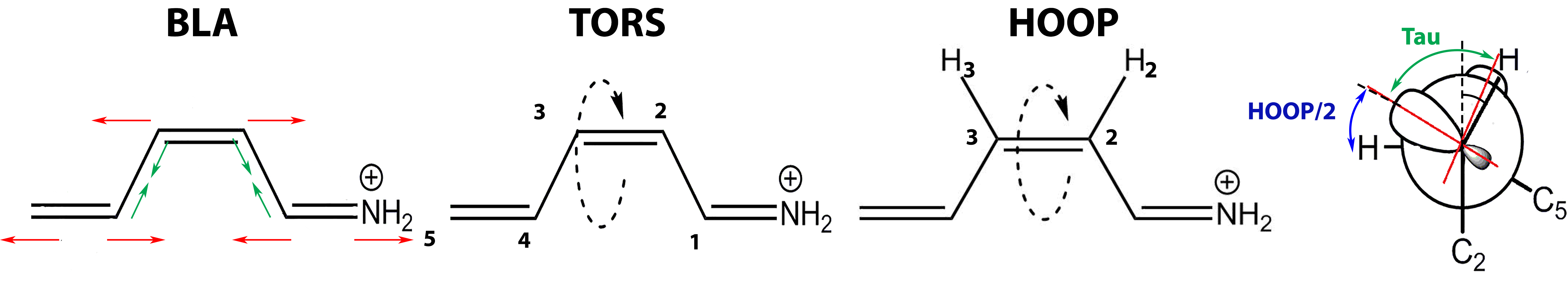}
    \caption{Schematic representation of BLA $r = \frac{d_{C_1C_2}+d_{C_3C_4}}{2} - \frac{d_{NC_1}+d_{C_2C_3}+d_{C_4C_5}}{3}$, Tors $\theta = \mathrm{dihedral}\left(\mathrm{C}_1\mathrm{C}_2\mathrm{C}_3\mathrm{C}_4\right)$, and HOOP $\phi = \mathrm{dihedral}\left(\mathrm{C}_1\mathrm{C}_2\mathrm{C}_3\mathrm{C}_4\right)- \mathrm{dihedral}\left(\mathrm{H}_2\mathrm{C}_2\mathrm{C}_3\mathrm{H}_3\right)$ (the relation between HOOP and Tau, shown on the right, is $\mathrm{Tau} = \mathrm{Tors}-\mathrm{HOOP}/2$) of \textit{cis}-PSB3. Reprinted with permission from Ref.[\citenum{marsili2019two}] (J. Chem. Theory Comput. 2020, 16, 10, 6032-6048). Copyright 2020 American Chemical Society.}
\label{F:modes}
\end{figure*}

The photoisomerization reaction from the cis to the trans isomer of PSB3 activated upon a vertical excitation from S$_0$ to S$_1$ is modeled employing the two-electronic-state three-vibrational-mode Hamiltonian derived in Ref.[\citenum{marsili2019two}] and refined in Ref.[\citenum{marsili2020quantum}]. In addition to previous models accounting for BLA and Tors,\cite{hahn2000quantum} this model introduces for the first time the HOOP coordinate and its coupling to Tors as essential elements to favor/prevent the formation of the trans photoproduct, as discussed in some detail in Ref.[\citenum{marsili2020quantum}] based on the analysis of trajectory-based dynamics. Figure~\ref{F:modes} schematically shows the three coordinates of the model, while we refer the interested reader to Ref.[\citenum{marsili2020quantum}] (Appendix A) to the details of the analytical model potentials.

The quantum dynamics (QD) of photoexcited \textit{cis}-PSB3 is simulated here for 200~fs using the Quantics code\cite{worth2020quantics} in the electronic diabatic basis (we recall that the model itself accounts for two electronic diabatic states of covalent and charge-transfer character\cite{marsili2019two}). The Short Iterative Lanczos (SIL) method is used, increasing the size of the basis until convergence is ensured. In this work, we only adopted the approximate kinetic energy operator of Ref.[\citenum{marsili2020quantum}] with a constant and diagonal metric tensor, whose components are $G^{rr}\equiv 1/ M_r=7.981 \cdot 10^{-5}$~a.u., $G^{\theta\theta}\equiv 1/I_{\theta}=2.599 \cdot 10^{-5}$~a.u., $G^{\phi\phi}\equiv 1/I_{\phi}= 40.375\cdot 10^{-5}$~a.u. As in Ref.[\citenum{marsili2020quantum}], the initial vibrational wavepacket (WP) is chosen to be a three-dimensional Gaussian centered at $r,\theta,\phi=0.1725~\textrm{a.u.},0,0$ with zero initial momentum and width $\sigma_r,\sigma_\theta,\sigma_\phi=0.07720,0.09165,0.20305$~a.u. along the three vibrational directions.

\begin{table}[!ht]
   \centering
   \caption{Definition of the primitive basis/grid used in this work; $N_{q}$ indicates the number of grid points and we give in parenthesis the values used in Ref.[\citenum{marsili2020quantum}]. $Q_{0}$, $\omega$ and $M$ are the  equilibrium position, frequency and mass for the harmonic oscillator (HO) basis functions (all values in a.u.), respectively. The column ``Grid'' indicates the domain spanned by Tors and HOOP.}
   \label{tab:hamiltonian_parameters}
   \begin{tabular}{c|cccccc}
   \hline\hline
   Mode     & Type     & $N_{q}$  & $Q_{0}$      &Grid & $\omega$ & $M$  \\ \hline
   BLA      & HO       & 30(32)   & 0.1725        & &0.005839575    & 14358   \\
   Tors     & Fourier  & 256(136)    & &$\pm \pi$ &  \\
   HOOP     & HO       & 60(60)     &  & $\pm \pi$ & \\ \hline
   \end{tabular}
   \label{T:basis-set}
\end{table}

Table~\ref{T:basis-set} reports information on the basis sets and on the grids that ensure fully converged QD results. As in Ref.[\citenum{marsili2020quantum}], harmonic oscillator (HO) basis functions are used along BLA and HOOP, while the Fourier basis is used for Tors. Note, in particular, that in Table~\ref{T:basis-set} we provide the number of grid points along each coordinate used in this work as well as the value used in Ref.[\citenum{marsili2020quantum}] (whose results will be reported in Section~\ref{sec: results}) for comparison. Furthermore, we compared the QD results obtained with Quantics to those generated with the ElVibRot code\cite{ElVibRot} used in Ref.[\citenum{marsili2020quantum}], to ensure convergence using the same primitive basis sets and with similar basis/grid sizes.

\begin{figure*}[ht!]
    \centering
    \includegraphics[width=1.0\textwidth]{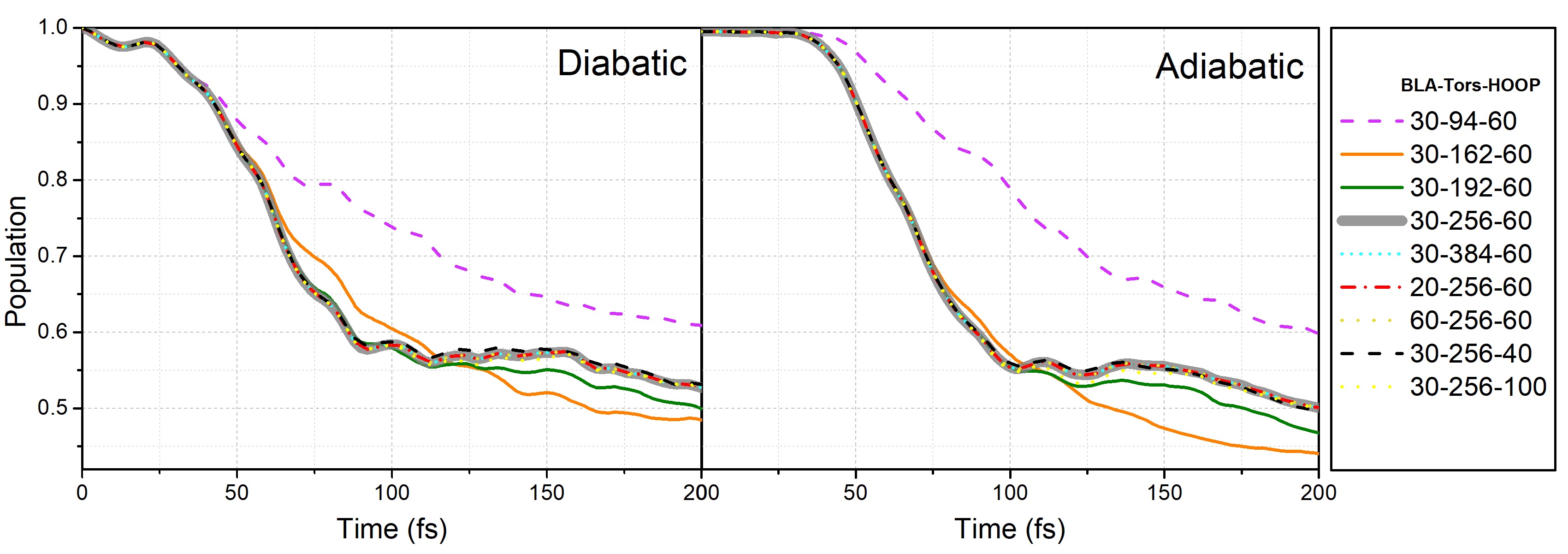}
    \caption{Time evolution of diabatic (left) and adiabatic (right) state populations, indicating the number of grid points  employed for BLA, Tors and HOOP, respectively.}
\label{F:tpop}
\end{figure*}

A key aspect of our work, which will be discussed in this section, is to highlight the need to perform extensive convergence analyses to ensure reliable QD results, while in Section~\ref{sec: results}, the focus will be put on the physical aspects of the quantum behavior of photoexcited \textit{cis}-PSB3 that our simulations unveil.

In Figure~\ref{F:tpop}, we report the time evolution of the population of the diabatic (left) and of the adiabatic (right) states that are initially fully populated. As indicated in the figure, the various curves refer to different combinations of the numbers of grid points, i.e., $N_q$ of Table~\ref{T:basis-set}, along the BLA, Tors ans HOOP coordinates. The main and critical observation that both diabatic and adiabatic populations allow us to point out is the strong dependence of the results on the grid along Tors. Our tests employed values of $N_q$ along Tors ranging from 94 to 384, thus we can conclude that with 256 grid points (thick gray curve in Figure~\ref{F:tpop}) convergence is achieved. Note that additional tests were carried out by increasing the primitive basis along the BLA (from 20 to 60) and the HOOP (from 40 to 100) modes, allowing us to confirm convergence with 30 and 60 grid points for BLA and HOOP, respectively, similarly to Ref.[\citenum{marsili2020quantum}]. 

\begin{figure*}[ht!]
    \centering
    \includegraphics[width=1.0\textwidth]{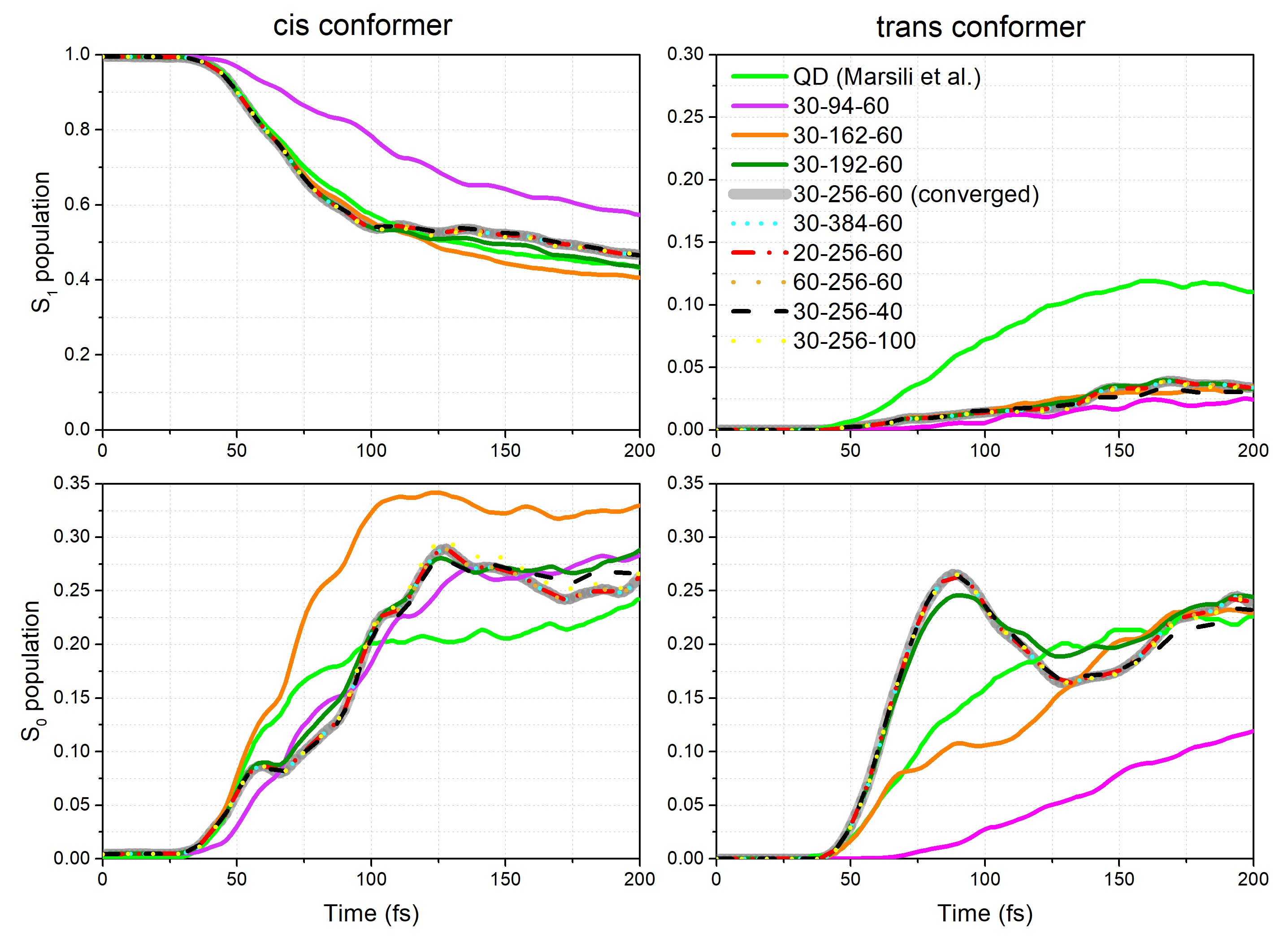}
    \caption{Time evolution of the adiabatic state populations, S$_0$ (lower panels) and S$_1$ (upper panels), for the cis (left panels) and trans (right panels) isomers. The various curves correspond to different combinations of the number of grid points; the light-green curves show to the results Ref.[\citenum{marsili2020quantum}], whereas the results that will be used for the analysis in Section~\ref{sec: results} are shown as thick gray lines.}
\label{F:ctpop}
\end{figure*}

Figure \ref{F:ctpop} shows the adiabatic populations of S$_0$ (lower panels) and S$_1$ (upper panels) as functions of time, resolved in the cis (left panels) and trans (right panels) isomers by integrating the nuclear probability densities in S$_0$ and in S$_1$ over all the BLA and HOOP domains but in the interval $-90^\circ<\textrm{Tors}<90^\circ$ for the cis isomer and outside of this interval for the trans isomer. In Figure \ref{F:ctpop}, we report for comparison the results of Ref.[\citenum{marsili2020quantum}] (light-green curves), that were obtained using $32-136-60$ grid points for BLA, Tors and HOOP, respectively. We note here that, while the behavior of the population in S$_1$ is qualitatively fairly stable with respect to the increase of the grid size, the S$_0$ populations of both isomers change qualitatively, and approach convergence only after a certain threshold (in the order magenta, light-green, orange, and dark-green, the curves correspond to 94, 136, 162, 192 and show a somehow non-monotonic convergence to the thick gray lines). In particular, the results suggest that when QD along Tors is not correctly described, the WP moving in the trans region is prevented from repopulating the cis isomer, thus missing the clear oscillations of the S$_0$ trans population with a period of $\sim$ 100 fs observed in the lower right panel of Figure~\ref{F:ctpop}. Similarly, the fast oscillations of the S$_0$ cis population (lower left panel) are non correctly reproduced.

The need for such a large primitive basis set suggests that the adoption of more efficient propagation methods than SIL, like Multiconfigurational Time Dependent Hartree  (MCTDH),\cite{meyer1990multi,beck2000multiconfiguration} might reveal particularly beneficial. Testing the performance of MCTDH in this three-dimensional photoisomerization model is also important since its better scalability with the number of coordinates (especially in the multilayer variant \cite{Vendrell2011,mlmctdh2}) makes it possibly applicable to study the QD of photoisomerizations involving a larger number of coordinates.

We performed extensive numerical tests, which are detailed in Appendix~\ref{app: mctdh}. There, we show that convergence to the exact numerical results is possible both in the single-set and in the multi-set formalism,\cite{beck2000multiconfiguration} the latter being more efficient in terms of computational time. The better performance of the multi-set ansatz can be rationalized due to the large differences between the ground and excited potential energy surfaces, which make specific single-particle functions (SPFs) more accurate in describing the WP. Interestingly, we demonstrate that achieving convergence requires a remarkably large number of SPFs, especially for the TORS coordinate in the ground state, whereas capturing the dynamics along BLA and HOOP is less demanding. Specifically, we used 160 and 110 SPFs along the Tors mode in the ground and in the excited state, respectively. An insufficient number of SPFs leads to notable inaccuracies in both diabatic and adiabatic populations after 100 fs and hinders the proper exchange of trans/cis  population observed in fully converged simulations (discussed below). This effect is also reflected in the computed quantum yield, emphasizing the need for careful convergence testing when using MCTDH for strongly anharmonic systems, especially if a large amount of kinetic energy accumulates in few degrees of freedom.

\section{Quantum and quantum-classical dynamics}\label{sec: results}
In this section, we discuss in detail the QD dynamics of photoexcited \textit{cis}-PSB3 in terms of the evolution of populations, quantum yield and vibrational kinetic energy. In addition, we assess again the performance of the quantum-classical trajectory-based methods used in Ref.[\citenum{marsili2020quantum}] in comparison with the converged QD results presented in Section~\ref{sec: model}.

Henceforth, we use the symbols $P^{\rm{S}_1}_{\text{cis}}$, $P^{\rm{S}_1}_{\text{trans}}$, $P^{\rm{S}_0}_{\text{cis}}$ and $P^{\rm{S}_0}_{\text{trans}}$ to indicate the time-dependent populations of the isomers cis in the excited state, trans in the excited state, cis in the ground and trans in the ground state, respectively.

\begin{figure*}[ht!]
    \centering
    \includegraphics[width=1.0\textwidth]{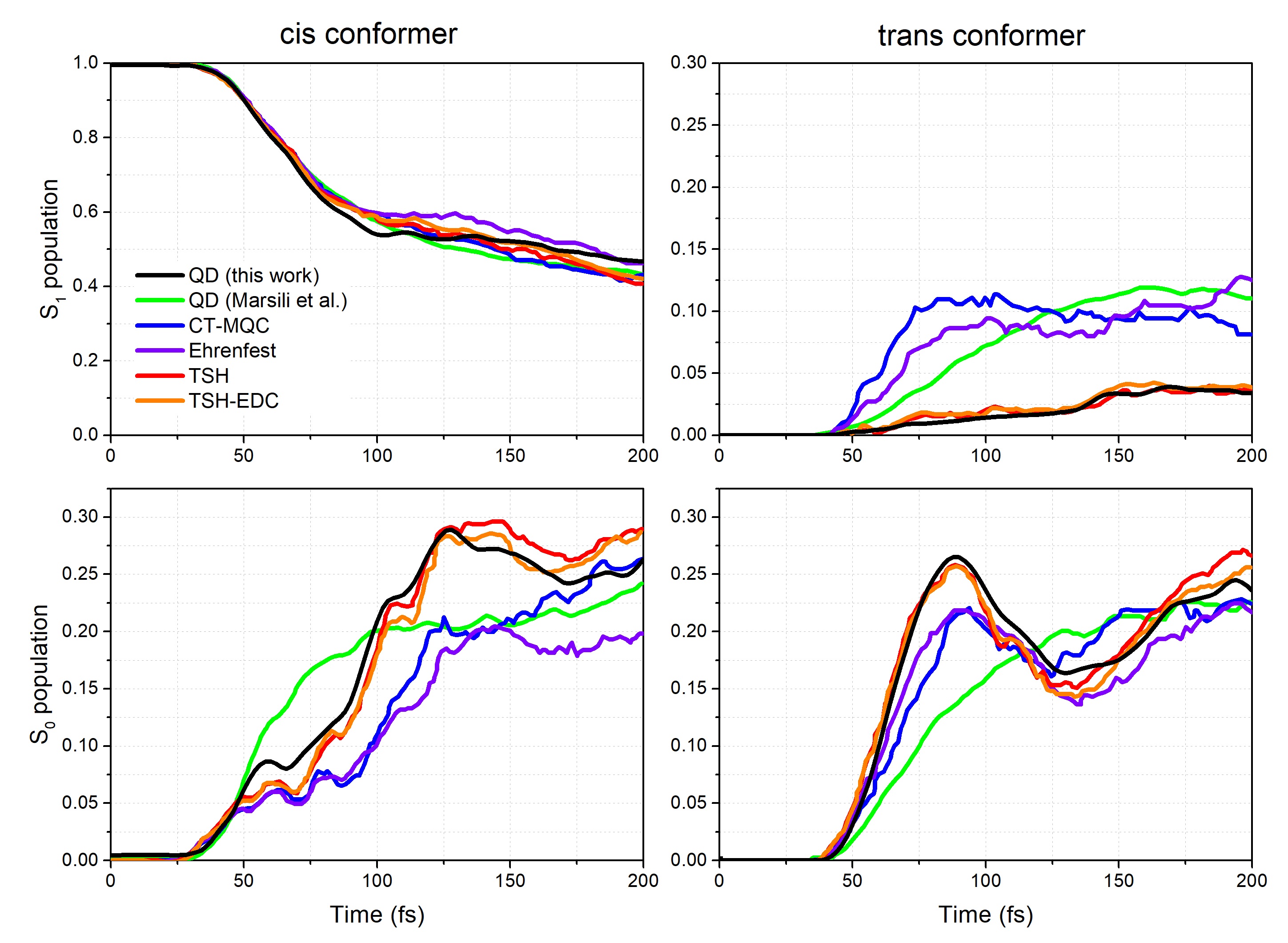}
    \caption{Time evolution of the populations of the cis (left panels) and trans (right panels) isomers in the ground state S$_{0}$ (lower panels) and in the excited state S$_{1}$ (upper panels). QD results of this work are shown in black, while QD results of Ref.[\citenum{marsili2020quantum}] are shown in light-green. Quantum-classical results with TSH (red), TSH-EDC (orange), Ehrenfest (purple) and CT-MQC (blue) are also shown.}
\label{F:pop}
\end{figure*}

Figure~\ref{F:pop} shows, similarly to Figure~\ref{F:ctpop} above, the evolution of $P^{S_1}_{\text{cis}}$ (upper left panel), $P^{S_1}_{\text{trans}}$ (upper right panel), $P^{S_0}_{\text{cis}}$ (lower left panel), $P^{S_0}_{\text{trans}}$ (lower right panel). The QD results of the present work are shown in black, along with the previous results of Ref.[\citenum{marsili2020quantum}] in light-green. Quantum-classical results have been obtained using the trajectory surface hopping method (TSH),\cite{Tully1990} also including energy-based decoherence (TSH-EDC) corrections,\cite{granucci2007} the multi-trajectory Ehrenfest method, and coupled-trajectory mixed quantum-classical (CT-MQC) algorithm.\cite{min2015} In Figure~\ref{F:pop}, red and orange lines indicate TSH and TSH-EDC, respectively, whereas purple lines correspond to Ehrenfest and blue lines to CT-MQC. Trajectory-based dynamics was performed in Ref.[\citenum{marsili2020quantum}] using the G-CTMQC code\cite{GCTMQC} interfaced with the QuantumModelLib potential library\cite{ModelLib} where the PSB3 model is implemented. 

As observed in Section~\ref{sec: model}, all simulations predict a similar behavior for $P^{S_1}_{\text{cis}}$. 
However, Figure~\ref{F:pop} clearly shows that the populations $P^{S_1}_{\text{trans}}$, $P^{S_0}_{\text{cis}}$ and $P^{S_0}_{\text{trans}}$
are substantially different from those reported in Ref.[\citenum{marsili2020quantum}], specifically: \\
(1) the excited-state trans population $P^{S_1}_{\text{trans}}$ is always remarkably smaller than the originally reported population; \\
(2) the ground-state cis population $P^{S_0}_{\text{cis}}$ is smaller up to $\sim$ 100 fs and then larger at later times, with the largest difference of $\sim$ 0.08 observed at $\sim$ 125 fs;\\
(3) the ground-state cis $P^{S_0}_{\text{cis}}$ and trans $P^{S_0}_{\text{trans}}$ populations show the most significant difference with respect to the original results, since the present QD calculations show two distinct  \textsl{oscillatory behaviors} that were completely absent in the previous results.

The QD results reported here demonstrate, as point (3) just highlighted, a qualitative different behavior of  $P^{S_0}_{\text{cis}}$ and $P^{S_0}_{\text{trans}}$ from the original results of Ref.[\citenum{marsili2020quantum}], which can be summarized as the appearance of two oscillatory patterns in S$_0$: the first is displayed by the cis population and starts at $\sim$ 60 fs, displaying several step-like increases up to $\sim$ 125 fs and with a period of $\sim$ 30 fs; the second is displayed by the trans population and shows a maximum at $\sim$ 85 fs, a minimum at $\sim$ 125 fs and a new increase up to $\sim$ 195 fs, while in Ref.[\citenum{marsili2020quantum}] a monotonic increase of the population was observed until approaching a plateau towards the end of the simulated dynamics.

The comparison between the converged QD results of the present work and the quantum-classical results of Ref.[\citenum{marsili2020quantum}] suggests that the conclusions on the performance of TSH/TSH-EDC and of Ehrenfest/CT-MQC need to be revisited. In particular, while a general agreement can be observed across methods, even though not always quantitative, we observe that surface-hopping-based methods predict the population dynamics in close agreement with QD simulations. Instead, Ehrenfest and CT-MQC lack quantitative agreement with the QD reference, in particular, by overestimating the excited-state trans population and conversely underestimating the ground-state populations. Note that, for this particular model, it is somehow expected that TSH and TSH-EDC results are very similar as well as Ehrenfest and CT-MQC. QD dynamics shows the emergence of a highly coherent behavior during the whole simulated dynamics, thus accounting for decoherence in TSH-EDC and CT-MQC does not affect the corresponding TSH or Ehrenfest dynamics (note that both electronic and nuclear evolution equations defining the CT-MQC algorithm have been cast such that they have the form of standard Ehrenfest equations to which coupled-trajectory terms are added to account for decoherence).

\begin{figure*}[ht!]
    \centering
    \includegraphics[width=0.7\textwidth]{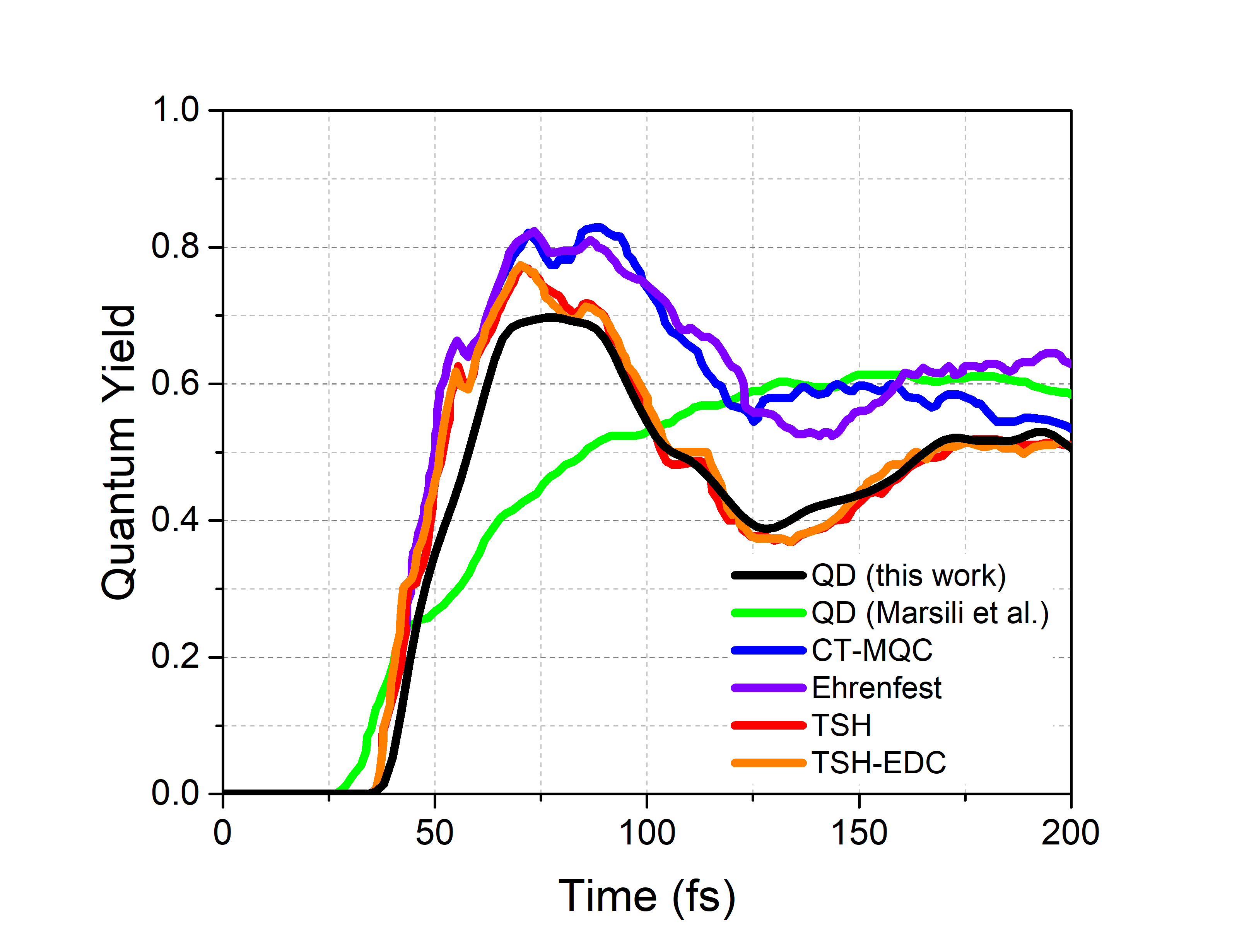}
    \caption{Time-dependent quantum yield obtained from the various simulations. The color code is the same as in Figure~\ref{F:pop}.}
\label{F:QY}
\end{figure*}

The time-dependent quantum yield of the photoreaction is determined here as the ratio of the trans products and all products, \cite{marsili2020quantum} namely as
\begin{equation}
QY = \frac{P^{\rm S_0}_{\text{trans}} + P^{\rm S_1}_{\text{trans}}}{P^{\rm S_0}_{\text{cis}} + P^{\rm S_0}_{\text{trans}} + P^{\rm S_1}_{\text{trans}}}
\end{equation}
and is shown in Figure~\ref{F:QY}. As expected, QD results of this work (black) manifest an oscillatory behavior which is completely missed by the QD results of Ref.[\citenum{marsili2020quantum}] (light-green). Instead, all quantum-classical methods correctly capture the appearance of the oscillations, with a period that is in excellent quantitative agreement with QD results. However, only the curves calculated based on TSH (red) and TSH-EDC (orange) almost perfectly follow the QD (black) curve along all the simulated dynamics. 

Particularly interesting is the larger value of the quantum yield reached at short times (about 0.75 at 75~fs according to the QD simulations of the present work) than the previous calculations of Ref.[\citenum{marsili2020quantum}]. Such a high quantum efficiency in the formation of the trans isomer, correctly captured by TSH and TSH-EDC, appears to be the consequence of a faster funneling process through the the conical intersection between S$_0$ and S$_1$, leading the WP (or the trajectories) to acquire a large amount of kinetic energy along the low-frequency Tors mode, with corresponding loss of potential energy as well as the damping of BLA and HOOP oscillations in their corresponding kinetic energies (see discussion below).

The oscillatory behavior of the quantum yield observed in Figure~\ref{F:QY} reflects the time evolution of the populations shown in Figure \ref{F:pop} and unveils a trans$\rightarrow$cis back-reaction occurring in the ground state in the $80~\textrm{fs} < t< 130~\textrm{fs}$ time window. Such a back-reaction appears clearly when analyzing in detail the behavior of the populations from 86 fs to 130 fs. In particular, in this time interval $P^{\rm S_0}_{\text{trans}}$ varies of $-0.1$, $P^{\rm S_0}_{\text{cis}}$ varies of $+0.16$, $P^{\rm S_1}_{\text{trans}}$ varies of $+0.01$, and, finally, $P^{\rm S_1}_{\text{cis}}$ varies of $-0.07$. Therefore, this analysis suggests that, in this time interval, $P^{\rm S_0}_{\text{cis}}$ increases mainly as a result of a decrease of $P^{\rm S_0}_{\text{trans}}$ (back-reaction), and only to a lesser extent of $P^{\rm S_1}_{\text{cis}}$ (internal conversion). Our hypothesis is that such a back-reaction is made possible due to the large amount of kinetic energy injected by the photoexcitation in the Tors coordinate when the system reaches S$_0$ (see below). 

As also done in Ref.[\citenum{marsili2020quantum}], we find instructive to analyze the time evolution of the kinetic energy along the BLA, Tors and HOOP coordinates, yielding information about the activation of the modes after the photoexcitation and about their effect on the oscillatory behavior of the populations.

\begin{figure*}[ht!]
    \centering
    \includegraphics[width=1.0\textwidth]{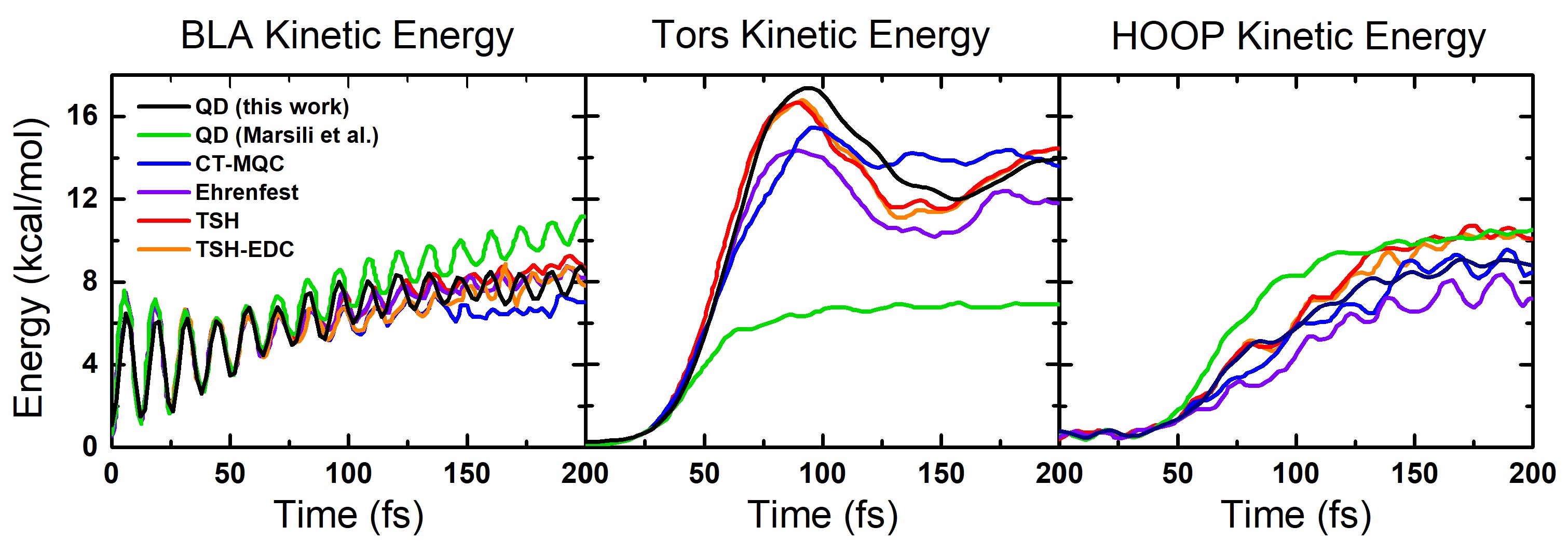}
    \caption{Time evolution of the nuclear kinetic energy along the BLA (left), Tors (center), and HOOP (right). The color code is the same used in Figure~\ref{F:pop}.}
\label{F:Kin}
\end{figure*}

Figure \ref{F:Kin} shows the time evolution of the expectation value of the kinetic energy along BLA (left), Tors (center) and HOOP (right). The color code is the same used in Figure~\ref{F:pop}. Once again, QD results obtained in this work are closely captured by the quantum-classical simulations, especially based on TSH and TSH-EDC, whereas significant discrepancies are observed with the original QD results of Ref.[\citenum{marsili2020quantum}].
Remarkably, the period of the HOOP kinetic energy oscillations matches the period of the step-like oscillatory behavior seen in the ground state \textit{cis}-PSB3 population indicating that the HOOP phase is capable of affecting the excited state decay of the cis isomer, as discussed in Ref.[\citenum{marsili2020quantum}] based on representative CT-MQC trajectories.
The largest deviation of our QD results from the original QD results of Ref.[\citenum{marsili2020quantum}] is observed for the Tors kinetic energy. The new findings suggest that the differences in the cis and trans populations stem from an underestimation of the Tors kinetic energy in the original simulations, due to the small size of the basis set used along this mode. Note that within the first 50 fs of dynamics all predictions are similar and they all show that the amplitude of the oscillations in the BLA kinetic energy is damped while the WP spreads toward the conical intersections at $\textrm{Tors} =\pm 90^\circ$. 
After 50 fs,  the quantum-classical methods  predict that the Tors motion acquires  kinetic energy which seems largely overestimated when compared to the original QD results,  with a maximum difference reached at 80 fs delay. Instead, the converged QD results agree very well with the quantum-classical ones.  The larger kinetic energy gained by Tors according to the new QD results is compensated by a smaller kinetic energy along the BLA and HOOP. A more detailed analysis suggests that, as observed above for the populations and the quantum yield, TSH and TSH-EDC predictions are in very good agreement with the QD reference as far as Tors is concerned, whereas, at least after 100 fs  CT-MQC gives a more accurate prediction for the HOOP kinetic energy. In addition, the loss in BLA kinetic energy and the damping of its oscillations predicted by the new QD is fairly well captured by the quantum-classical methods.
Such features are likely due to a more effective decoherence in the BLA direction, due to the larger WP spreading along Tors and HOOP coordinates (as will be discussed below). 
As discussed above, the large kinetic energy available along the Tors coordinate is responsible for the trans$\rightarrow$cis back-reaction in S$_0$ and, therefore, for the oscillatory behavior of the ground-state trans population and of the quantum yield.  

Finally, we analyze the actual QD in a reduced two-dimensional configuration space using the coordinates Tors and HOOP. The reduced density, obtained by integrating over the BLA domain, is shown in Figure~\ref{F:WPhoop} at different times, as indicated in the figure, and is resolved in the S$_0$ (left panels) and S$_1$ (right panels) contributions. The two-dimensional HOOP-Tors reduced density shows a larger spreading in S$_0$ than the one predicted in the original QD calculations, a difference which can be ascribed to the larger basis used along Tors to achieve convergence. 
Interestingly, the regions explored in Figure \ref{F:WPhoop} by the reduced density in S$_0$ seem to agree with those visited by the CT-MQC trajectories (see Figure 5 of Ref.[\citenum{marsili2020quantum}]).

\begin{figure*}[ht!]
    \centering
    \includegraphics[width=0.75\textwidth]{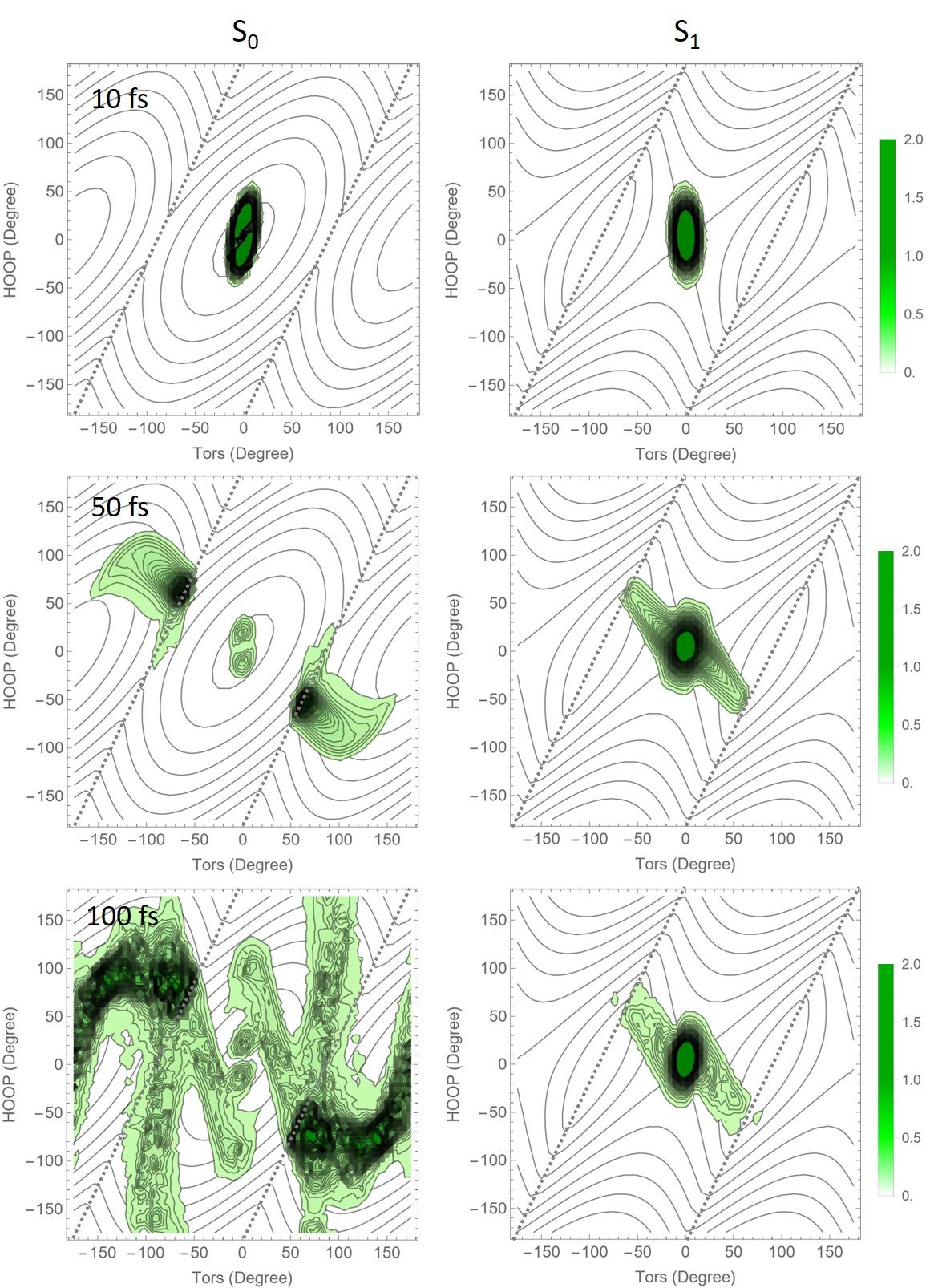}
    \caption{Two-dimensional Tors-HOOP nuclear density at 10~fs (top panels), 50~fs (middle panels) and 100~fs (bottom panels), resolved in the S$_0$ (left panels) and S$_1$ (right panels) contributions. Adiabatic S$_0$ (left) and S$_1$ (right) potential energy surfaces are represented via gray isocontours for BLA = 0; the dashed lines indicate the S$_0$/S$_1$ intersection seam.} 
\label{F:WPhoop}
\end{figure*}

Overall, we conclude that larger basis sets than those previously employed in Ref.[\citenum{marsili2020quantum}] are necessary to achieve converged QD of the photoisomerization of \textit{cis}-PSB3 using the model of Refs.[\citenum{marsili2019two, marsili2020quantum}]. Converged QD results show that quantum-classical methods correctly capture this photoreaction along all the simulated dynamics. TSH and TSH-EDC show quantitative agreement for a large class of properties, while Ehrenfest and CT-MQC capture only qualitatively the QD behavior.

\section{Conclusions}\label{sec: conclusions}
In this work, we presented an in-depth analysis of the quantum dynamics initiated by photoexcitation in the two-state three-dimensional model for the isomerization of \textit{cis}-PSB3 proposed in Refs.[\citenum{marsili2019two, marsili2020quantum}]. In particular, we revisited the numerical results and the conclusions of Ref.[\citenum{marsili2020quantum}] based on new and reliable calculations providing the quantum wavepacket dynamics, the time evolution of the electronic populations, and photoisomerization quantum yield.

Photoisomerizations play a key role in many chemical and biological processes and, when occurring in the ultrafast subpicosecond regime, their mechanism is usually characterized by the existence of conical intersections and nonadiabatic couplings between electronic states. They are, therefore, an intrinsically quantum-mechanical process whose natural description is obtained through quantum dynamics. However, quantum dynamics simulations are often time consuming or even prohibitive, making trajectory-based methods a very attractive alternative. Additionally, the application of trajectory-based methods, once properly validated, may offer the possibility to unveil details of the reaction mechanism that are not easily accessible in a quantum dynamics framework. This was exactly the case of the analysis of the photoisomerization in \textit{cis}-PSB3,\cite{marsili2020quantum} where the trajectories predict, even in a dramatically reduced model chromophore, the existence of a phase relationship between Tors and HOOP velocities whose signs at the time of the transition from S$_1$ to S$_0$ determine the success of the isomerization in a scotopic biological photoreceptor \cite{Weingart2008,Schapiro2011,Weigart2011,Schnedermann2018}. These considerations highlight the need to provide the community with reliable reference quantum dynamics data for the model of Refs.[\citenum{marsili2019two, marsili2020quantum}]. The present work demonstrates an excellent agreement between quantum dynamics and surface-hopping dynamics, an observation that has been revisited here with respect to Ref.[\citenum{marsili2020quantum}].

Our new computations also provide a rationale to understand why reaching full convergence of quantum dynamics can become very challenging even in reduced-dimensionality models of photoisomerizations. In the present study, capturing the large amplitude motion along Tors and HOOP in the ground state and the large amount of kinetic energy acquired by Tors after the nonadiabatic transition from S$_1$ to S$_0$ requires large basis sets. Since both features are a natural consequence of the large excitation energy provided to the system by photoexcitation and of the small number of coordinates that can dissipate it, we believe that these results are of general interest for quantum dynamics studies of photoisomerization reactions.

Concerning the information on the photoisomerization reaction extracted from the quantum dynamics, the results presented in this work suggest the appearance of a first coherent oscillatory behavior in the early-time relaxation towards the electronic ground state forming the cis isomer. The initial appearance of population of the cis isomer in S$_0$ is about 50~fs and the oscillations in the population agree with the oscillations of the kinetic energy along the HOOP mode, thus, these results seem to suggest that HOOP plays a key role in guiding the relaxation after photoexcitation and the isomerization reaction, as was observed for the same model from the analysis of representative classical trajectories in Ref.[\citenum{marsili2020quantum}] and also previously when reporting on the multi-scale modelling of the dynamics of the entire biological photoreceptor \cite{Weingart2008,Schapiro2011,Weigart2011,Schnedermann2018}. 

As stressed also in Ref.[\citenum{marsili2020quantum}], the model does not take into consideration either additional (internal) vibrational modes or the environment, and due to the reactant planarity, population transfer from S$_{1}$ to S$_{0}$ takes place symmetrically along the clockwise and counterclockwise directions Tors $>$ 0 and Tors $<$ 0. Furthermore, the energy pumped into the system via the initial electronic excitation cannot be dissipated, causing the nuclear wavepacket to remain in S$_{1}$, as it is evident in all the numerical results present in this work. 
Our analysis of the quantum dynamics suggests that the majority of the population of the initially populated state decays initially to S$_{0}$ in the trans isomer, reaching a maximum at 80 fs, after which a second coherent oscillatory feature drives a cis population increase in S$_0$ up to $\sim$125 fs, which is accompanied by a decrease of the trans isomer  in S$_0$. This implies, as discussed above, the trans$\rightarrow$cis back-reaction in the ground state. After 130 fs, the cis population in S$_0$ decays in favor of the trans population in S$_0$. Specifically, these results are evidence of the occurrence of a trans/cis interconversion in the ground state, since at times larger than $80$ fs the cis and trans populations on S$_1$ undergo only mild changes, respectively decreasing and increasing by $\sim$ 0.05 up to 200 fs.

\section*{Acknowledgements}
MA, MG and FS gratefully acknowledge financial support from the U.S. Department of Energy, Office of Science, Office of Basic Energy Sciences, Chemical Sciences, Geosciences and Biosciences Division under award no. DE-SC0022225. FS and MG  acknowledge  support from the CRESCENDO project, PRIN: PROGETTI DI RICERCA DI RILEVANTE INTERESSE NAZIONALE-Bando 2022, Prot. 2022HL9PRP. 
M.O. is grateful for partial support provided by EU funding within the MUR PNRR ``National Center for Gene Therapy and Drugs based on RNA Technology'' (Project no. CN00000041 CN3 RNA) - Spoke 6. M.O. is also grateful to the NSF CHE-SDM A for Grant No. 2102619 and to the MUR for a PRIN 2022 grant No. 2022K3AY2K.
FA acknowledges financial support from the French Agence Nationale de la Recherche under the grants No. ANR-20-CE29-0014 (ANR Q-DeLight project) and No. ANR-23-ERCC-0002 (ANR STROM project).

\appendix
\section{MCTDH simulations}\label{app: mctdh}
Although the numerically exact solution of the time-dependent Schr\"odinger equation on a full direct-product grid offers a highly accurate representation of the wavepacket (WP), its dimensionality, and consequently the computational cost, increases exponentially with the number of degrees of freedom. This exponential scaling limits the applicability of exact quantum propagation to relatively small systems, typically with fewer than 5–6 degrees of freedom. Thanks to the impressive methodological progress of Multi-Configurational Time-Dependent Hartree (MCTDH)\cite{beck2000multiconfiguration} and its multi-layer (ML) extension (ML-MCTDH),\cite{Vendrell2011,mlmctdh2} handling systems with a large number of degrees of freedom (typically up to 20-30 degrees of freedom with MCTDH and up to hundreds with its ML extension) is now feasible. MCTDH exploits a variationally optimized, time-dependent basis to represent the WP. In practice, the WP is represented as the sum of products of time-dependent single-particle functions (SPFs). Therefore, the accuracy of the dynamics in MCTDH depends not only on the number of primitive basis functions used for each degree of freedom, but also, and strongly, on the number of SPFs.

\begin{figure*}[ht!]
    \centering
\includegraphics[width=1.0\textwidth]{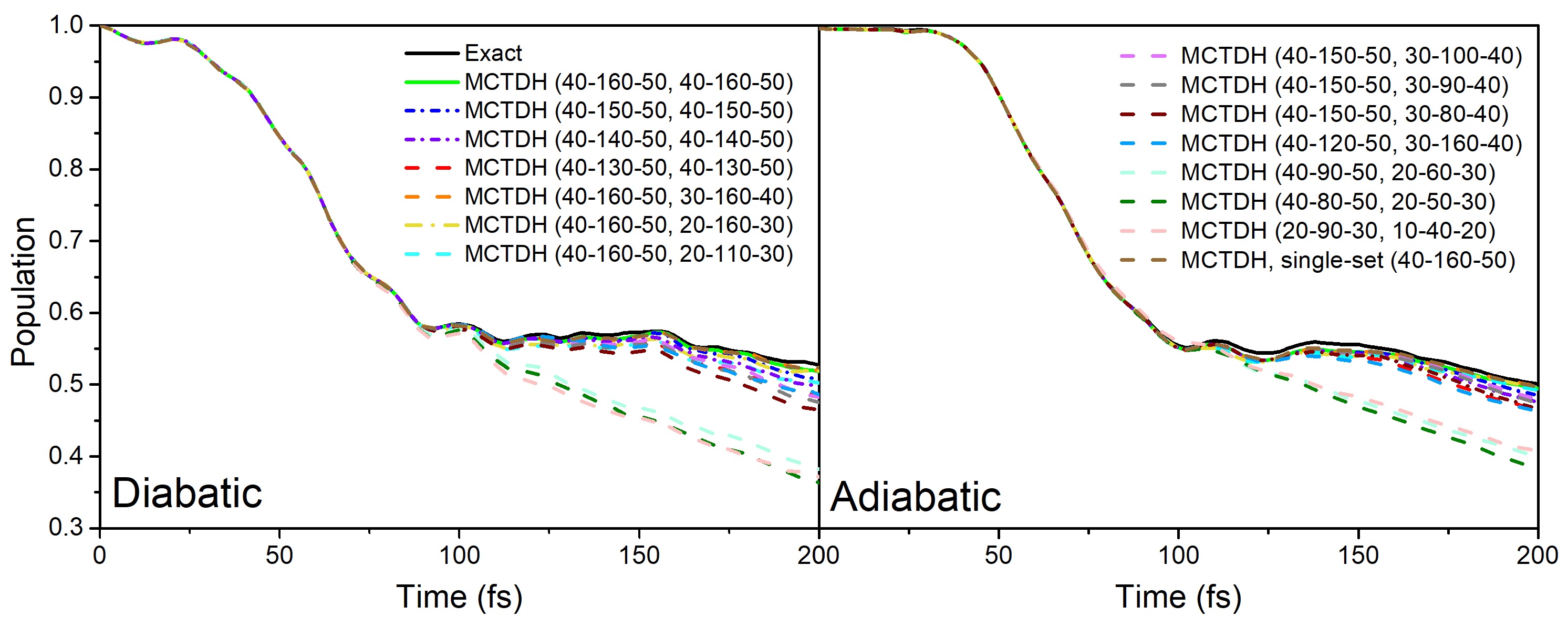}
    \caption{Dependence of the time evolution of diabatic (left) and adiabatic (right) state populations on the number of SPFs assigned to each mode with MCTDH. The values in parentheses indicate the number of SPFs used for the BLA, Tors and HOOP, respectively, for the ground state, followed by those for the excited state. The results are compared with our numerically exact propagation (black line) presented in Section~\ref{sec: model}. In all simulations, the BLA, Tors, and HOOP modes were represented using 80, 256, and 90 grid points, respectively.} 
\label{F:totpop}
\end{figure*}
In this Appendix, we systematically assess the convergence of the quantum dynamics simulation performed with MCTDH on the three-dimensional potential model describing the photoisomerization of PSB3. In Figure~\ref{F:totpop}, we report the time evolution of the population of the diabatic (left) and
of the adiabatic (right) states. As indicated in the figure, the various curves refer to different combinations of the numbers of SPFs adopted for BLA, Tors, and HOOP coordinates. We employed the multi-set formalism, which allows assigning different sets of SPFs to each vibrational mode on different electronic states. For comparison, we also performed the dynamics with the single-set formalism for a specific case, indicating that the multi-set formalism yields results equivalent to the single-set approach when using the same number of SPFs for both states. Moreover, for MCTDH simulations, we slightly increased the number of grid points for BLA and HOOP to provide more flexibility in assigning different numbers of SPFs. As shown in Figure~\ref{F:totpop}, achieving convergence requires a large number of SPFs (up to 160) for the Tors mode in the ground state, whereas approximately 110 SPFs are sufficient in the excited state. This behavior is expected, given the significant amount of kinetic energy acquired by the Tors mode in the ground state. It is also evident that the dynamics is not particularly sensitive to the number of SPFs assigned to the HOOP and BLA modes, as seen by comparing the light-green solid line with the orange and yellow lines, where the number of SPFs for BLA and HOOP in the excited state was progressively reduced. A significant deviation from the reference results (in black), consistently observed after 100 fs, occurs when a reduced number of SPFs is used for Tors, particularly in the ground state. 

\begin{figure*}[ht!]
    \centering
\includegraphics[width=1.0\textwidth]{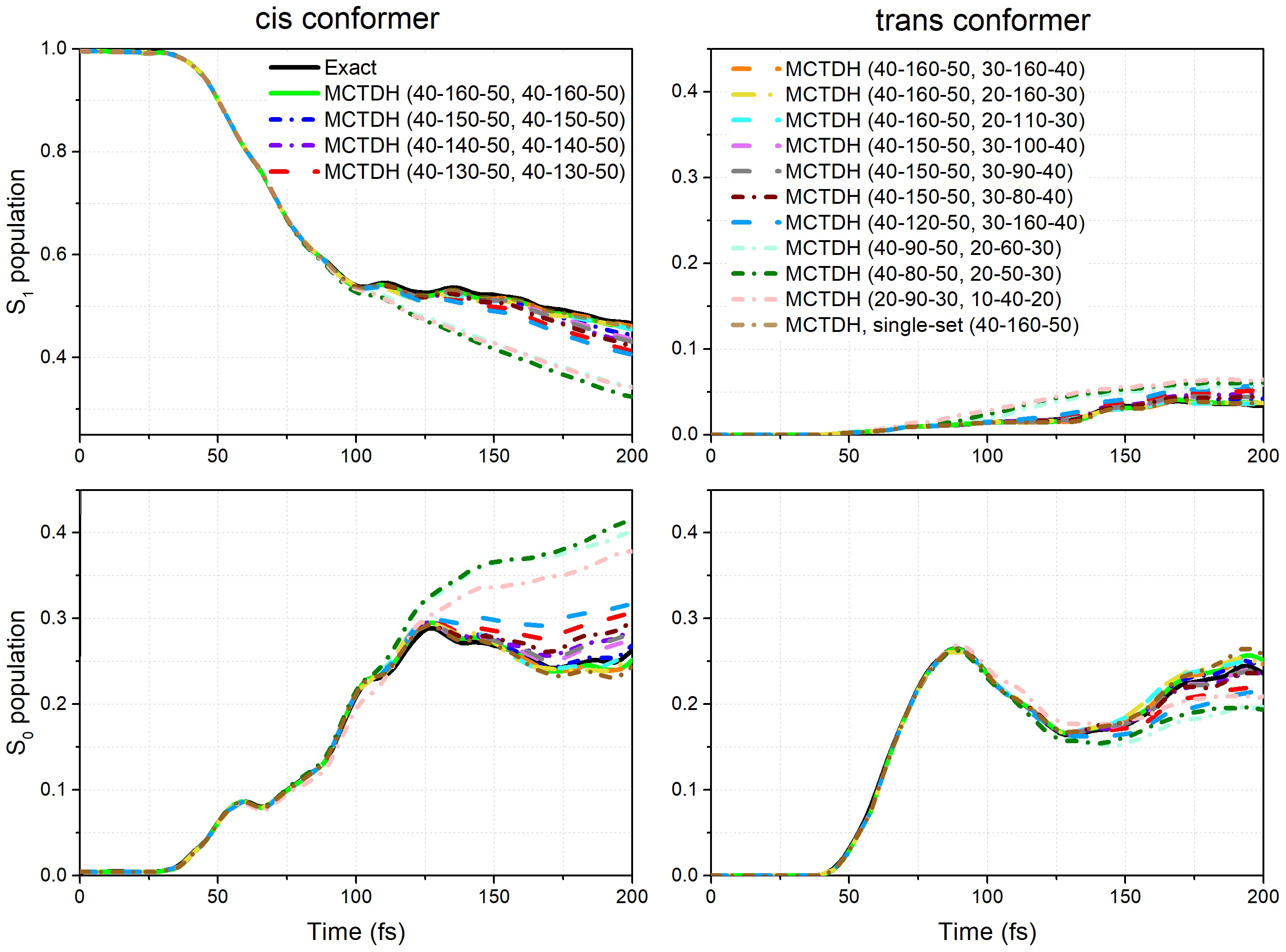}
    \caption{Dependence of time evolution of the populations of the cis (left panels) and trans (right panels) isomers in the ground state S$_{0}$ (lower panels) and in the excited state S$_{1}$ (upper panels) on the number of SPFs assigned to each mode, with MCTDH WP.  The values in parentheses indicate the number of SPFs used for the BLA, Torsion, and HOOP modes, respectively, for the ground state, followed by those for the excited state. The results are compared with a numerically exact propagation (black line). In all simulations, the BLA, Torsion, and HOOP modes were represented using 80, 256, and 90 grid points, respectively.} 
\label{F:WPhoop}
\end{figure*}
\begin{figure*}[ht!]
    \centering
\includegraphics[width=0.9\textwidth]{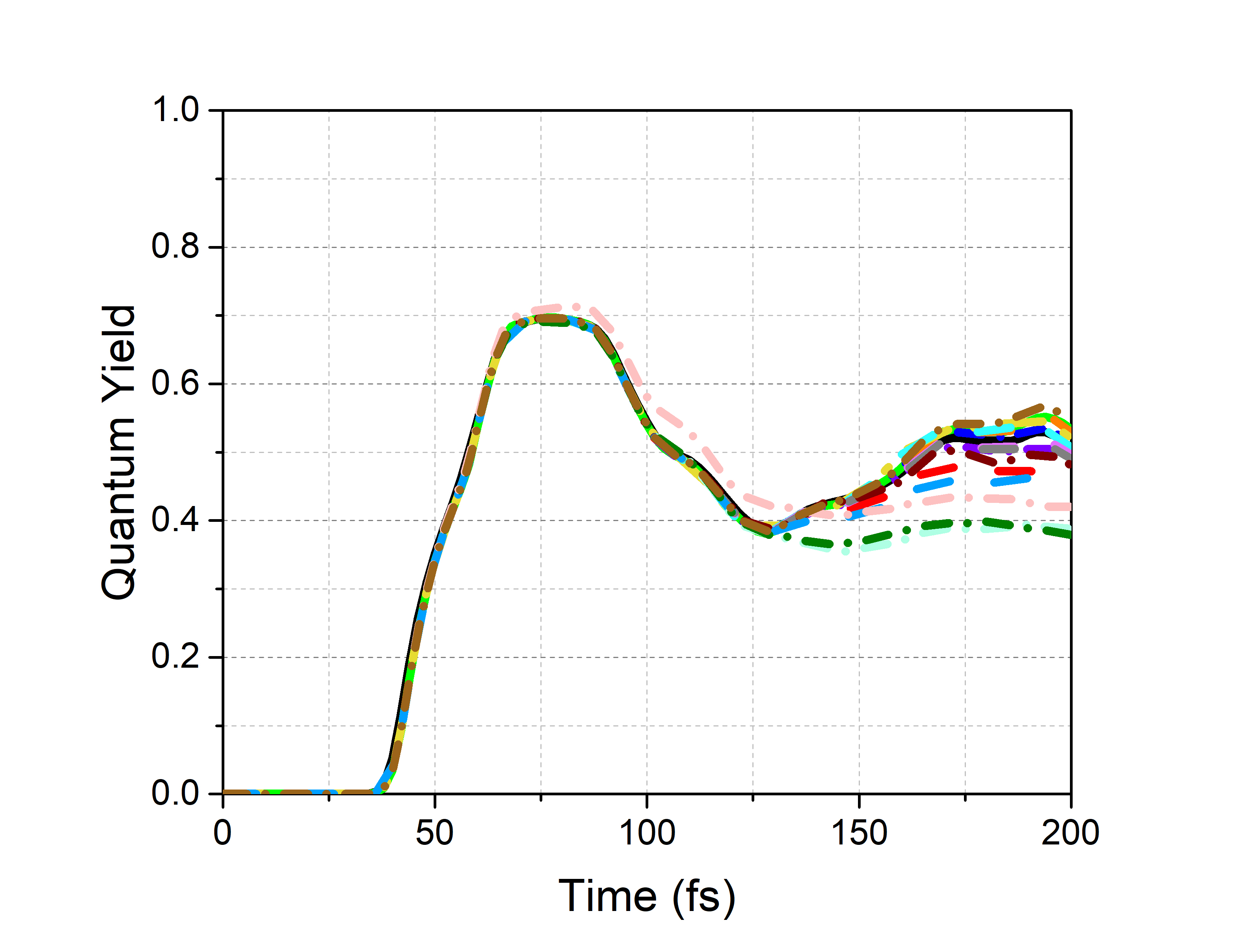}
    \caption{Similar to Figure~\ref{F:WPhoop} but for the time-dependent quantum yield. The color code is the same as in Figure~\ref{F:WPhoop}.} 
\label{F:QY app}
\end{figure*}

Figure~\ref{F:WPhoop} shows, similarly to Figure~\ref{F:ctpop}, the time evolution of the populations $P^{\rm{S}_1}_{\text{cis}}$ (upper left panel), $P^{\rm{S}_1}_{\text{trans}}$ (upper right panel), $P^{\rm{S}_0}_{\text{cis}}$ (lower left panel) and $P^{\rm{S}_0}_{\text{trans}}$ (lower right panel) calculated using MCTDH with different numbers of SPFs assigned to each mode on different electronic states. Interestingly, the S$_1$ population for both isomers appears to be more sensitive to the basis size (i.e., the number of SPFs) than to the primitive grid size, as evidenced by the significant deviation from the reference observed after 100 fs. A similar trend is observed for the S$_0$ population, with the cis isomer showing a strong dependence on the number of SPFs in both the S$_0$ and S$_1$ states. Specifically, an insufficient number of SPFs appears to hinder the cis-to-trans population transfer in S$_0$ observed in the fully converged quantum dynamics simulation after 130 fs. This effect is also reflected in the time-dependent quantum yields shown in Figure~\ref{F:QY app}, where the overestimated cis population in S$_0$ after 130 fs, caused by insufficient convergence, leads to a reduced quantum yield at long times.

\providecommand{\latin}[1]{#1}
\makeatletter
\providecommand{\doi}
  {\begingroup\let\do\@makeother\dospecials
  \catcode`\{=1 \catcode`\}=2 \doi@aux}
\providecommand{\doi@aux}[1]{\endgroup\texttt{#1}}
\makeatother
\providecommand*\mcitethebibliography{\thebibliography}
\csname @ifundefined\endcsname{endmcitethebibliography}
  {\let\endmcitethebibliography\endthebibliography}{}

\end{document}